\documentclass[sigconf]{acmart}

\pdfoutput=1

\setcopyright{rightsretained}
\usepackage{enumitem}
\usepackage{bm}
\usepackage{xspace}
\usepackage{color}
\usepackage{mathtools}
\usepackage{subcaption}
\usepackage{booktabs}
\usepackage{multirow}
\usepackage{adjustbox}
\usepackage[ruled, vlined]{algorithm2e}

\setlist[itemize]{leftmargin=18pt}
\setlist[enumerate]{leftmargin=24pt}
\newcommand{\vpara}[1]{\vspace{0.08in}\noindent\textbf{#1 }}
\newcommand{\hide}[1]{} 
\newcommand{\numberthis}{\addtocounter{equation}{1}\tag{\theequation}}
\newcommand\hmm[1]{\ifnum\ifhmode\spacefactor\else2000\fi>1000 \uppercase{#1}\else#1\fi}

\newcommand{\eg}{{\sl e.g.\xspace}}
\newcommand{\ie}{{\sl i.e.\xspace}}
\newcommand{\etc}{{\sl etc.}}

\newcommand{\wrt}{{\sl w.r.t.\xspace}}

\newcommand{\trans}{^\top}
\newcommand{\obj}{\mathcal{O}}
\newcommand{\ttl}{\mathcal{L}}  
\newcommand{\nodeset}{\mathcal{V}}
\newcommand{\pairset}{\mathcal{S}}
\newcommand{\expdist}[1]{\mathrm{Exp}\left( #1 \right)}
\newcommand{\dirdist}[2]{\mathrm{Dir}_{#1}\left( #2 \right)}
\newcommand{\gammadist}[1]{\Gamma\left( #1 \right)}
\newcommand{\mb}{\mathbf}
\newcommand{\eulere}{\operatorname{e}}

\newcommand{\bmeta}{\bm{\eta}}
\newcommand{\bmrho}{\bm{\rho}}
\newcommand{\bmtau}{\bm{\tau}}
\newcommand{\bmphi}{\bm{\phi}}
\newcommand{\bmP}{\bm{P}}
\newcommand{\bmPhi}{\bm{\Phi}}
\newcommand{\bmTheta}{\bm{\Theta}}

\newcommand{\pa}{\textit{Mordo}\xspace}
\newcommand{\pb}{\textit{Wong}\xspace}

\newcommand{\pd}{\textit{Stephen}\xspace}

\newcommand{\nodevis}{\textit{node visibility}\xspace}
\newcommand{\pathsel}{\textit{path selectivity}\xspace}
\newcommand{\crosssyn}{\textit{cross-meta-path synergy}\xspace}

\begin{document}
\title{PReP: Path-Based Relevance from a Probabilistic Perspective in Heterogeneous Information Networks}

\author{Yu Shi}
\affiliation{\institution{University of Illinois at Urbana-Champaign}}
\email{yushi2@illinois.edu}
\author{Po-Wei Chan}
\affiliation{\institution{University of Illinois at Urbana-Champaign}}
\email{pchan12@illinois.edu}
\author{Honglei Zhuang}
\affiliation{\institution{University of Illinois at Urbana-Champaign}}
\email{hzhuang3@illinois.edu}
\author{Huan Gui}
\affiliation{\institution{University of Illinois at Urbana-Champaign}}
\email{huangui2@illinois.edu}
\author{Jiawei Han}
\affiliation{\institution{University of Illinois at Urbana-Champaign}}
\email{hanj@illinois.edu}


\begin{abstract}
As a powerful representation paradigm for networked and multi-typed data, the heterogeneous information network (HIN) is ubiquitous.
Meanwhile, defining proper relevance measures has always been a fundamental problem and of great pragmatic importance for network mining tasks.
Inspired by our probabilistic interpretation of existing path-based relevance measures, we propose to study HIN relevance from a probabilistic perspective.
We also identify, from real-world data, and propose to model \crosssyn, which is a characteristic important for defining path-based HIN relevance and has not been modeled by existing methods.
A generative model is established to derive a novel path-based relevance measure, which is data-driven and tailored for each HIN.
We develop an inference algorithm to find the maximum a posteriori (MAP) estimate of the model parameters, which entails non-trivial tricks.
Experiments on two real-world datasets demonstrate the effectiveness of the proposed model and relevance measure.
\end{abstract}

\copyrightyear{2017} 
\acmYear{2017} 
\setcopyright{acmlicensed}
\acmConference{KDD '17}{}{August 13--17, 2017, Halifax, NS, Canada}
\acmPrice{15.00}
\acmDOI{http://dx.doi.org/10.1145/3097983.3097990}
\acmISBN{978-1-4503-4887-4/17/08}

{
%
%


}



\maketitle

\section{Introduction}\label{sec::intro}
In real-world applications, objects of various types are often interconnected with each other.
These objects, together with their relationship, form numerous heterogeneous information networks (HINs) \cite{shi2017survey, sun2013mining}.
Bibliographical information network is a typical example, where researchers, papers, organizations, and publication venues are interrelated.
A fundamental problem in HIN analysis is to define proper measures to characterize the relevance between node pairs in the network, 
which also benefits various downstream applications, such as similarity search, recommendation, and community detection \cite{shi2017survey, sun2013mining}.

Most existing studies derive their HIN relevance measures on the basis of \textit{meta-path} \cite{shi2017survey, sun2013mining, sun2011pathsim}, 
which is defined as a concatenation of multiple node types linked by corresponding edge types.
Based on the concept of \textit{meta-path}, researchers have proposed PathCount, PathSim \cite{sun2011pathsim}, and path constrained random walk \cite{lao2010relational} to measure relevance between node pairs.
On top of these studies, people have explored the ideas of incorporating richer information \cite{he2014exploiting, yao2014pathsimext} and more complex typed structures \cite{fang2016semantic, huang2016meta, shi2014hetesim} to define more effective relevance scoring functions, or adding supervision to derive task-specific relevance measures \cite{chen2017task, wang2011learning, yu2012user}.

\vpara{The probabilistic perspective.}
While building upon this powerful \textit{meta-path} paradigm, we aim to additionally understand and model relevance from the probabilistic point of view. 
In this regard, we establish a probabilistic interpretation of existing HIN relevance measures, which is achieved by modeling the generating process of all path instances in an HIN and deriving the relevance of a node pair from the likelihood of observing the path instances connecting them.
Relevance and likelihood can be connected by this approach because only a small portion of node pairs in an HIN are actually relevant; and a proper generating process has low likelihood to generate the path instances between each of these relevant node pairs.
We will detailedly discuss this probabilistic interpretation in Sec.~\ref{sec::proba_view}.
Moreover, as a starting point for studying HIN relevance from the probabilistic perspective, we focus the scope of this paper on the basic \textit{unsupervised} scenario.
Meanwhile, we assume that the meta-paths of interest are already given.
That is, we defer the study on the cases with label information and meta-path selection to future work.

In order to determine relevance between any pair of nodes, we have the key insight that a path-based HIN relevance should contain three characteristics -- \nodevis, \pathsel, and \crosssyn{} -- which we describe in the following paragraphs.

\vpara{Node visibility.} 
One straightforward way to derive relevance in an HIN is PathCount \cite{sun2011pathsim}. 
For a meta-path $t \in \{1, \ldots, T\}$, PathCount is defined as the number of paths, $P_{st}$ or equivalently $P_{\langle uv \rangle t}$, under this meta-path between a node pair $s = (u, v) \in \nodeset \times \nodeset$, \ie, $\mathit{PathCount}^{(t)}(u, v) \coloneqq P_{\langle uv \rangle t}$.
One obvious drawback of this measure is that it favors nodes with high \nodevis, \ie, nodes with a large number of paths.
To resolve this problem, \cite{sun2011pathsim} proposed to penalize PathCount by the arithmetic mean of the numbers of cycles attached to the two involved nodes, \ie, 
$
\mathit{PathSim}^{(t)}(u,v) \coloneqq \frac{2 \cdot P_{\langle uv \rangle t}}{P_{\langle uu \rangle t} + P_{\langle vv \rangle t}}.
$
A similar design to model \nodevis can be found in JoinSim \cite{xiong2015top}, which is defined as PathCount penalized by geometric mean of the cycle numbers.

\begin{figure}[t]
  \centering
  \begin{subfigure}[m]{0.45\linewidth}
    \centering\includegraphics[width=\linewidth]{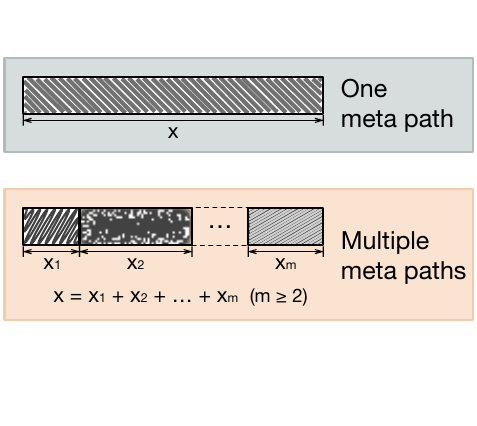}
    \caption{\label{fig::comp_score_one}}
  \end{subfigure}
  \begin{subfigure}[m]{0.54\linewidth}
    \centering\includegraphics[width=\linewidth]{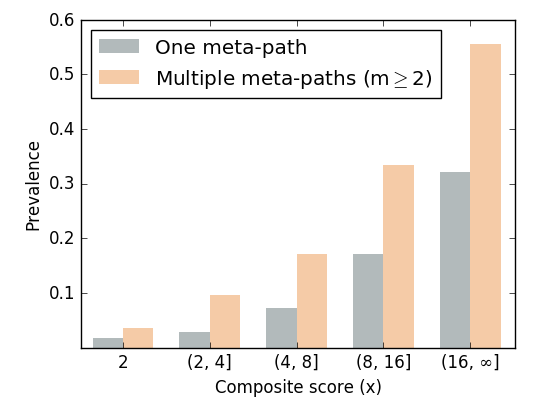}
    \caption{\label{fig::nonlinearity_bar_graph}}
  \end{subfigure}
  \caption{(\subref{fig::comp_score_one}) The same composite score ($x$) may be aggregated from different number of meta-paths, where score is represented by the length of the rectangles and each fill pattern represents a meta-path. (\subref{fig::nonlinearity_bar_graph}) An observation made from an entity resolution task on the DBLP dataset that if linear combination is used to compute the composite score, node pairs with paths under multiple meta-paths are more likely to be relevant than those under only one meta-path. Prevalence is defined as the number of relevant node pairs divided by the total number of node pairs.}\label{fig::nonlinearity_in_data}
\end{figure}

\vpara{Path selectivity.} 
Given any method defining relevance score under one meta-path, a natural question is how to combine multiple meta-paths to derive a unified relevance score -- henceforth referred to as the \textit{composite score}.
To achieve this goal, Sun et al. \cite{sun2011pathsim} proposed to assign different weights to different meta-paths, and compute the composite score via linear combination.
Let $\mb{w} = \{w_1, \ldots, w_T\}$ with $w_t$ being the weight for meta-path $t$, the composite score of PathCount is given by $\mathit{PathCount}_{\mb{w}}(u, v) \coloneqq \sum_{t=1}^T w_t \cdot \mathit{PathCount}^{(t)}(u, v)$.
Similarly, one can define $\mathit{PathSim}_{\mb{w}}(u, v)$.
This linear combination approach is adopted by follow-up works with multiple applications \cite{shi2017survey, sun2013mining}, including personalized entity recommendation problem \cite{yu2014personalized}, outlier detection  \cite{kuck2015query, zhuang2014mining}, \etc{} 
The weights assigned or inferred in these cases specify how selective each meta-path is. 
The larger the \pathsel, the more significant this meta-path is in contributing to the composite score.

\vpara{Cross-meta-path synergy.}
Suppose linear combination is used to find the composite score as in the previous paragraph, 
the two scenarios shown in Fig.~\ref{fig::comp_score_one} would receive the same composite score ($x$), where $x_i$ equals to the score from the $i$-th meta-path multiplied by the corresponding weight.
However, we have the observation that, when meta-paths do not clearly correlate, the latter scenario tends to imply a higher relevance. 
We take an entity resolution task on the DBLP dataset as example, which aims to merge author mentions that refer to the same entity.
In this task, each node stands for an author mention, and each meta-path represents that two author mentions have both published papers in one particular research area.
We label two author mentions as relevant if and only if they refer to the same entity, and we use PathCount with uniform weights as an example to compute the composite score.
Results presented in Fig.~\ref{fig::nonlinearity_in_data} shows that with the same composite score, node pairs associated by paths under multiple meta-paths are more likely to be relevant than those under only one meta-path.
We refer to this phenomenon as \crosssyn.
We interpret this phenomenon as given the occurrence of one path, the happenstance of another path under the same meta-path may not be surprising, while the co-occurrence of two paths under two uncorrelated meta-paths may be a strong signal of relevance.
Moreover, we should also realize that not necessarily all meta-path pairs are uncorrelated, which has been observed in a special type of HIN \cite{shi2016dynamics}.
This implies \crosssyn does not necessarily exist between all pairs of meta-paths, and we deem a good relevance measure should reflect this difference.

\vpara{Challenges and contributions.}
Regarding the three pivotal characteristics for path-based HIN relevance discussed above, the major challenge lies in how to integrate all these characteristics in a unified framework.
We tackle this challenge by studying path-based relevance from a probabilistic perspective, and deriving relevance measure from a generative model. 
Since the model parameters are trained to fit each HIN, the derived relevance measure enjoys the property of being data-driven. 
That is, the derived relevance measure is tailored for each HIN.
Lastly, we summarize our contributions as follows:
\begin{enumerate}
\item
We establish the probabilistic interpretation of existing path-based HIN relevance measures. 
\item
We identify and propose to model \crosssyn, an important characteristic in path-based HIN relevance.
\item
We propose a novel relevance measure based on a generative model, which is data-driven and tailored for each HIN, and develop an inference algorithm with non-trivial tricks.
\item
Experiments on two real-world HINs corroborate the effectiveness of our proposed model and relevance measure.
\end{enumerate}

\section{Prelimineries}\label{sec::prelim}
In this section, we introduce the concepts and notations used in this paper.

\begin{definition}[Heterogeneous Information Network]
An \textbf{information network} is a directed graph $G = (\mathcal{V}, \mathcal{E})$ with a node type mapping $f : \mathcal{V} \rightarrow \mathcal{A}$ and an edge type mapping $g: \mathcal{E} \rightarrow \mathcal{R}$. Particularly, when the number of node types $|\mathcal{A}| > 1$ or the number of edge types $|\mathcal{R}| > 1$, the network is called a \textbf{heterogeneous information network} (HIN).
\end{definition}

Due to the typed essence of HINs, paths that associate node pairs can be grouped under different meta-paths. We formally define meta-paths as follows.

\begin{definition}[Meta-Path]
A \textbf{meta-path} is a concatenation of multiple nodes or node types linked by edge types.
\end{definition}

An example of a meta-path is $[\mathrm{author}] \xrightarrow{\mathrm{writes}} [\mathrm{paper}] \xrightarrow{\mathrm{writes}^{-1}} [\mathrm{author}]$, where a phrase in the brackets represents a node type and a phrase above the arrow refers to an edge type. 
When context is clear, we simply write [author]--[paper]--[author]. 
In this paper, we study the relevance problem when a set of meta-paths of interest is predefined by users.

To ease presentation, we focus on unweighted HINs, and model path count defined as follows.
Note that the path-based model to be proposed in this paper can be extended to the weighted case. 

\begin{definition}[Path Count]
The \textbf{path count} of a meta-path $t \in \{1, \ldots, T\}$ between a node pair $s = (u, v) \in \nodeset \times \nodeset$ is the number of concrete path instances under this meta-path that start from node $u$ to node $v$, which is denoted by $P_{st}$ or $P_{\langle uv \rangle t}$.
\end{definition}

Note that the relevance score given by the PathCount measure \cite{sun2011pathsim} is exactly the path count of a meta-path between a node pair.

Lastly, we introduce the probability distributions to be used. 
\begin{definition}
The probability density functions of three probability distributions used in this paper are given as follows.
\begin{enumerate}
\item        
Exponential distribution $\expdist{\tilde\lambda}$ with rate parameter $\tilde\lambda > 0$:
$$
p(x) = \tilde\lambda \eulere^{\tilde\lambda x} \quad (x > 0).
$$
\item
Gamma distribution $\gammadist{\tilde\alpha, \tilde\beta}$ with shape parameter $\tilde\alpha > 0$ and rate parameter $\tilde\beta > 0$:
$$
p(x) = \frac{\tilde\beta^{\tilde\alpha}}{\Gamma(\tilde\alpha)} x^{\tilde\alpha - 1} \eulere^{-\tilde\beta x} \quad (x > 0),
$$
where $\Gamma(\tilde\alpha) = \int_{0}^{\infty} t^{\tilde\alpha - 1} \eulere^{-t} dt$ is the gamma function.
\item
Symmetric Dirichlet distribution $\dirdist{L}{\tilde\alpha}$ of order $L$ and concentration parameter $\tilde\alpha$:
$$
\qquad \; \; p(x_1, \ldots, x_L) = \frac{\gammadist{\tilde\alpha L}}{\gammadist{\tilde\alpha}^L} \prod_{i=1}^L x_{i}^{\tilde\alpha-1} \quad (x_i > 0 \; \mathrm{and} \; \sum_{i=1}^L x_{i} = 1),
$$
where $\Gamma(\cdot)$ is the gamma function. 
\end{enumerate}
\end{definition}

We denote $\expdist{x \; ; \; \tilde\lambda} \coloneqq p(x)$ the probability density function of $\expdist{\tilde\lambda}$, and denote $x \sim \expdist{\tilde\lambda}$ if $x$ is generated from $\expdist{\tilde\lambda}$.
Similar notations are also used for $\gammadist{\tilde\alpha, \tilde\beta}$ and $\dirdist{L}{\tilde\alpha}$.

\section{Probabilistic Interpretation of Existing Relevance Measures}\label{sec::proba_view}
In this section, we illustrate the probabilistic interpretation of existing path-based HIN relevance measures.
We achieve this by studying the generating process of path counts between node pairs in an HIN, which contains a connection between relevance and the negative log likelihood.
Suppose the path count under meta-path $t$ between node pair $s$ is generated from an exponential distribution
$$
P_{st} \sim \expdist{\lambda}, 
$$
with fixed rate $\lambda$, then in terms of the rank it yields, the negative log likelihood of all observed paths under meta-path $t$ between node pair $s$ will be equivalent to the PathCount under meta-path $t$
\begin{align*}
-\mathit{LL}^{(t)}(s) &= - \log (\lambda \eulere^{- \lambda P_{st}}) = \lambda P_{st} - \log \lambda \\
& \propto P_{st} + \mathit{const} = \mathit{PathCount}^{(t)}(s) + \mathit{const}.
\end{align*}

Further, if we assume path instances under different meta-paths are generated from exponential distribution with meta-path-specific rates $\mb{w} = (w_1, w_2,  \ldots, w_T)$, \ie, $P_{st} \sim \expdist{w_t}$, then the negative log likelihood of all observed path counts will be equivalent to PathCount with weights $\mb{w}$ for linear combination
\begin{align*}
-\mathit{LL}(s) &= - \log ( \prod_t w_t \eulere^{- w_t P_{st}}) = \sum_t w_t P_{st} - \sum_t \log w_t \\
& = \sum_t w_t P_{st} + \mathit{const} = \mathit{PathCount}_{\mb{w}}(s) + \mathit{const}.
\end{align*}

Moreover, if we assume each node pair $s$ has pair-specific generating rate proportional to a parameter $\kappa_s$, \ie, $P_{st} \sim \expdist{{w_t}/{\kappa_s}}$, then the negative log likelihood of observed path counts will be
$
-\mathit{LL}(s) = \sum_t w_t \cdot \frac{P_{st}}{\kappa_s} + T \log \kappa_s + \mathit{const}.
$
For node pair $s=(u,v)$, if we drop the logarithm term and set $\kappa_s$ to be the arithmetic mean of the cycle count of the involved nodes $u$ and $v$, the formula becomes
$$
\sum_t w_t \cdot \frac{2 \cdot P_{\langle uv \rangle t}}{P_{\langle uu \rangle t} + P_{\langle v v \rangle t}} = \mathit{PathSim}_{\mb{w}}(s)
$$
which is identical to PathSim with weights $\mb{w}$ for linear combination. In lieu of arithmetic mean, if we set $\kappa_s$ to be the geometric mean of the same quantities, we get 
$
\sum_t w_t \cdot \frac{P_{\langle uv \rangle t}}{\sqrt{P_{\langle uu \rangle t} \cdot P_{\langle vv \rangle t}}},
$
which is identical to JoinSim with weights $\mb{w}$ for linear combination.
Note that all the relevance measures discussed in this section are special cases of our relevance measure to be proposed in the next section.

\section{Proposed Model and Relevance}\label{sec::model}
With the relevance--likelihood connection established in Sec.~\ref{sec::proba_view}, we propose our \textbf{P}ath-based \textbf{Re}levance from \textbf{P}robabilistic perspective (PReP) likewise by modeling the generating process of path counts between node pairs, and further aim to model the three important characteristics.
In a nutshell, the proposed generative-model-based relevance measure consists of two major parts:
(i) inferring model parameters by finding the maximum a posteriori (MAP) estimate to fit the input HIN, and
(ii) deriving relevance score between any node pair based on the learned model.

\begin{table}[t!]
\centering
\resizebox{0.48\textwidth}{!}{
\begin{tabular}{| c | c |}
\hline
\textbf{Symbol}  & \textbf{Definition} \\ 
\hline \hline
$\nodeset$ & The set of all nodes\\
$\pairset$ & The set of all nontrivial node pairs\\
$T \in \mathbb{N}$ & The number of meta-paths\\
$K \in \mathbb{N}$ & The number of generating patterns\\
\multirow{2}{*}{$\bm{P} \in \mathbb{R}^{|\pairset| \times T}$} & The observed path counts between node pairs\\ 
& over each meta-path\\
\hline
$\bm{\eta} \in \mathbb{R}^{T}$ & The \pathsel\\
$\bm{\tau} \in \mathbb{R}^{|\pairset|}$ & The \textit{node pair visibility}\\
$\bmrho \in \mathbb{R}^{|\nodeset|}$ & The \nodevis\\
$\bmTheta \in \mathbb{R}^{|\pairset| \times K}$ & The generating patterns over meta-paths \\ 
$\bmPhi \in \mathbb{R}^{K \times T}$ & The choices of generating patterns between node pairs\\
\hline
$\alpha \in \mathbb{R}_{+}$ & The shape parameter of the gamma prior\\
$\beta \in (0, 1)$ & The concentration parameter of the Dirichlet prior\\
\hline
\end{tabular}
}
\caption{Summary of symbols}\label{tab::symbol}
\end{table}

\begin{figure*}[t]
 \centering\includegraphics[width=\linewidth]{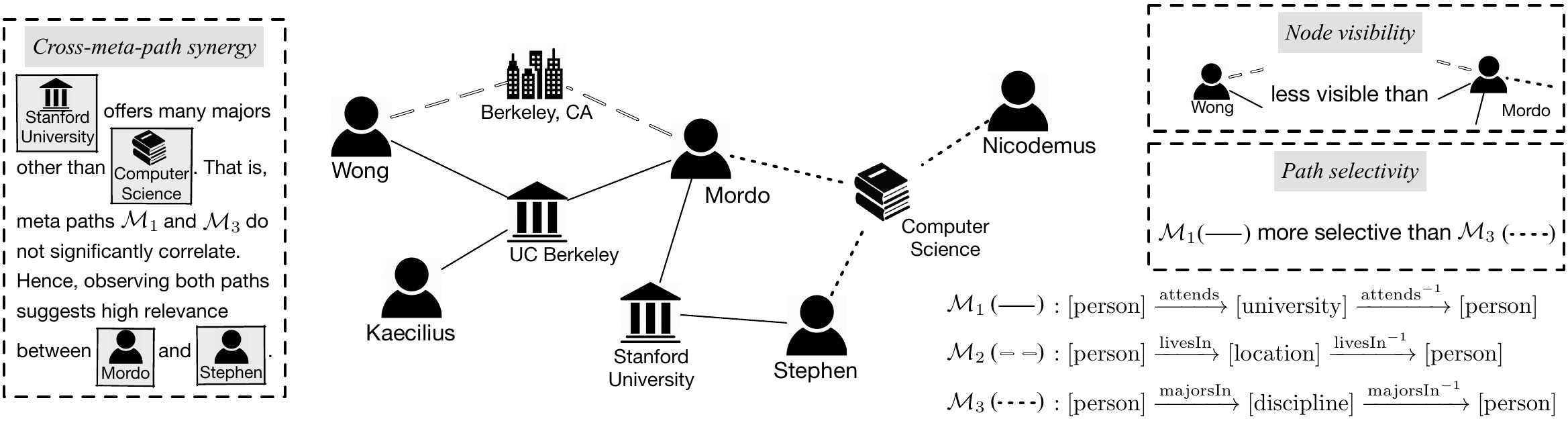}
 \caption{Toy example for one part of an HIN, consisting of four node types: person, university, location, and discipline.}\label{fig::toy_network}
\end{figure*}

\subsection{The PReP Model}\label{sec::path_gen}
Following the existing HIN relevance measures discussed in Sec.~\ref{sec::proba_view}, we assume the path count, $P_{st}$ or $P_{\langle uv \rangle t}$, between node pair $s = (u, v)$ under meta-path $t$ is generated from an exponential distribution with rate $\lambda_{st}$, \ie, $P_{st} \sim \expdist{\lambda_{st}}$.
To capture \nodevis, \pathsel, and \crosssyn, we must design $\lambda_{st}$ in a way that can model these three characteristics.

According to the property of exponential distribution, if a random variable $X$ is generated from $\expdist{\tilde\lambda}$, then the expectation of $X$ will be $1 / \tilde\lambda$.
Bearing this in mind, we introduce three components to model the three characteristics as follows.
\begin{itemize}
\item
Both the \nodevis of $u$ and that of $v$ affect the generation of path instances. We consider the visibility of this pair of node as \textit{node pair visibility}, $\tau_s$, which is positively correlated with the expectation of $P_{st}$.
\item
We let path instances under the same meta-path share the same \pathsel. Denote $\eta_t$ the \pathsel for meta-path $t$. $\eta_t$ is negatively correlated with the expectation of $P_{st}$.
\item
Each node pair with paths in between can be linked by path instances under a different set of meta-paths.
We assume an underlying \textit{meta-path distribution} $\bm{\psi}_{s} = [\psi_{s1}, \ldots, \psi_{sT}]$ for node pair $s$, where $\sum_{t=1}^T \psi_{st} = 1$ and $\psi_{st} \geq 0$. 
As a distribution over meta-paths, $\bm{\psi}_{s}$ models the semantics of the relevance between this node pair, because each meta-path carries its own semantic meaning.
With further design to be introduced, $\bm{\psi}_{s}$ also serves as the basis to capture \crosssyn.
$\psi_{st}$ is positively correlated with the expectation of $P_{st}$.
\end{itemize}
Putting the above three components together considering their correlation with the expectation of $P_{st}$, we find path count generating process as
\begin{equation}\label{eq::path_gen}
P_{st} \sim \expdist{\frac{\eta_t}{\tau_s \psi_{st}}},
\end{equation}
where the detailed illustration and design of the three components are to be further discussed in this section. Note that while we only discuss unweighted HINs in this paper, the use of exponential distribution in Eq.~\eqref{eq::path_gen} enables the model to handle weighted HINs, where paths are associated with real-valued path strengths, and $P_{st}$ may not be integers to reflect the path strengths.

Since node pairs with no paths under any predefined meta-path should trivially receive the lowest possible relevance score, we only model the generation of path counts between node pairs with paths in between -- henceforth referred to as nontrivial node pairs -- and we denote $\pairset$ the set of all nontrivial node pairs.

\vpara{Illustrative example.}
To better illustrate how each component design affects the path generation process, we present a toy example in Fig.~\ref{fig::toy_network}, which shows a part of an HIN with four node types: person, university, location, and discipline. 
We concern three meta-paths in this network: $\mathcal{M}_1: [\mathrm{person}] \xrightarrow{\mathrm{attends}} [\mathrm{university}] \xrightarrow{\mathrm{attends}^{-1}} [\mathrm{person}]$, $\mathcal{M}_2: [\mathrm{person}] \xrightarrow{\mathrm{livesIn}} [\mathrm{location}] \xrightarrow{\mathrm{livesIn}^{-1}} [\mathrm{person}]$, $\mathcal{M}_3: [\mathrm{person}] \xrightarrow{\mathrm{majorsIn}} [\mathrm{discipline}] \xrightarrow{\mathrm{majorsIn}^{-1}} [\mathrm{person}]$.

\vpara{Decoupling node pair visibility.}
To model \textit{\nodevis}, we decouple node pair visibility $\tau_s$ in Eq.~\eqref{eq::path_gen} into two parts as in PathSim and JoinSim discussed in Sec.~\ref{sec::proba_view}. 
The two parts correspond to the \nodevis $\rho_u$ and $\rho_v$, respectively, where $s = (u, v)$, and $\rho_z > 0$ for all $z \in \nodeset$. 
In our design, we let 
\begin{equation}\label{eq::decoupling}
\tau_{(u, v)} = \rho_u \rho_v
\end{equation}
as in JoinSim because decoupling by multiplication eases model inference, which will be made clear in the next paragraph.

Since a trivial rescaling -- multiplying all $\rho_z$ by a constant and multiplying all $\eta_t$ by the square of the same constant -- leads to exactly the same model (Eq.~\eqref{eq::path_gen}), we further regularize $\rho_z$ by a gamma prior with a constant rate parameter
\begin{equation}\label{eq::rho_prior}
\rho_z \sim \gammadist{\alpha, 1}.
\end{equation}
Note that we arbitrarily set the rate parameter to be $1$ since the shape of the distribution is solely determined by the shape parameter $\alpha$. 
We choose gamma distribution as the prior for $\rho_z$ because it is the conjugate prior for the exponential distribution,
and this fact will largely facilitate the inference algorithm as we will show in Sec.~\ref{sec::alg}. 
To determine the shape parameter $\alpha$, we fit the gamma distribution to the total path count each node has, $\{\sum_{t=1}^T \sum_{\tilde{z} \in \nodeset} P_{\langle z \tilde{z} \rangle t}\}_{z \in \nodeset}$, in the HIN as a rough prior information.

\vpara{Path selectivity at meta-path level.}
We assume path instances under meta-path $t$ share the same \pathsel $\eta_t$.
In the scope of this paper, where supervision is not available, we assume uninformative prior on $\eta_t$.
In future work where supervision is provided, we can further learn $\eta_t$ by minimizing the difference between supervision and model output to derive a task-specific relevance measure.

\vpara{Cross-meta-path synergy and generating patterns.}
As discussed in Sec.~\ref{sec::intro}, we have observed the existence of \crosssyn in real-world HIN, and this characteristic has not been modeled by existing HIN relevance measures.
In case meta-paths do not correlate, we may simply add a Dirichlet prior, with concentration parameter smaller than $1$, over meta-path distribution $\bm{\psi}_s$ for all node pair $s$.
This use of Dirichlet prior resembles latent Dirichlet allocation (LDA) \cite{blei2003latent}, where the Dirichlet prior prefers sparse distributions, \ie, most entries of $\bm{\psi}_s$ tend to be $0$.
Therefore, the co-occurrence of paths under different meta-paths gets a lower likelihood from this prior, and attains a higher relevance score under our relevance--likelihood connection.

However, in reality, it would not be surprising to see two people attending UC Berkeley also both live in the City of Berkeley.
This implies \crosssyn does not necessarily exist between all pairs of meta-paths, \eg, it may not exist between meta-path $\mathcal{M}_1$ and meta-path $\mathcal{M}_2$ in the toy example of Fig.~\ref{fig::toy_network}. 
To address this situation, we introduce a new component -- \textit{generating patterns}.
Each of a total of $K$ generating patterns is a distribution over the $T$ meta-paths, where meta-paths that often co-occur between node pairs will also be included in a common generating pattern, and when a node pair $s$ generates a path instance in between, it would first choose generating pattern $k$ with probability $\phi_{sk}$, and then choose meta-path $t$ from this generating pattern with probability $\theta_{kt}$.
Formally, we describe this process as
\begin{equation}\label{eq::gen_pat}
\psi_{st} = \sum_{k=1}^K \phi_{sk} \theta_{kt},
\end{equation}
where $\bm{\phi}_{s} = [\phi_{s1}, \ldots, \phi_{sK}]$ is node pair $s$'s choices of generating patterns, such that $\sum_{k=1}^K \phi_{sk} = 1$, $\phi_{sk} \geq 0$; and $\bm{\theta}_{k} = [\theta_{k1}, \ldots, \theta_{kT}]$ is generating pattern $k$'s distribution over meta-paths, such that $\sum_{t=1}^T \theta_{kt} = 1$, $\theta_{kt} \geq 0$.

A symmetric Dirichlet prior is then enforced on $\bmphi_s$, so that synergy will be recognized between and only between meta-paths from different generating patterns 
\begin{equation}\label{eq::phi_prior}
\bmphi_s \sim \dirdist{K}{\beta},
\end{equation}
where $\beta \in (0, 1)$ is the concentration hyperparameter. 

With this design, our model gives a lower likelihood and higher relevance score to \pa and \pd (same university, same major) than \pa and \pb (attending UC Berkeley and living in the City of Berkeley) in the toy example of Fig.~\ref{fig::toy_network} by learning a generating pattern that includes both $\mathcal{M}_1$ and $\mathcal{M}_2$. 
Whereas, other relevance measures cannot achieve this desired relationship as presented in Tab.~\ref{tab::linearity}, unless we set the weights $w_2 > w_3$, or equivalently assert $\mathcal{M}_2$ (location) is always less selective than $\mathcal{M}_3$ (discipline).

\begin{table}[t!]
\centering
\resizebox{0.48\textwidth}{!}{
\begin{tabular}{l | c | c | c | c | c | c}
\toprule
Measure                             & Node Pair & $\mathcal{M}_1$ & $\mathcal{M}_2$ & $\mathcal{M}_3$ & Composite  & Truth \\ \midrule
\multirow{2}{*}{PathCount} &  \pa \& \pb  &        $1$         &        $1$        &        $0$        &  $w_1 + w_2$  & $-$   \\ \cline{2-7}
                                           &  \pa \& \pd  &        $1$         &        $0$        &        $1$        &  $w_1 + w_3$  & $+$  \\ \hline
\multirow{2}{*}{PathSim}    &  \pa \& \pb  &        $0.67$         &         $1$        &       $0$         &  $0.67 w_1 + w_2$  & $-$   \\ \cline{2-7}
                                           &  \pa \& \pd  &        $0.67$         &         $0$        &       $1$         &  $0.67 w_1 + w_3$  & $+$   \\ \hline
\multirow{2}{*}{RWR ($C = 0.9$)}          &  \pa \& \pb  &        $0.29$         &         $0.47$        &       $0$         &  $0.29 w_1 + 0.47w_2$  & $-$   \\ \cline{2-7}
                                           &  \pa \& \pd  &        $0.25$         &         $0$        &       $0.31$         &  $0.25 w_1 + 0.31w_3$  & $+$   \\ \hline \hline
\multirow{2}{*}{PReP}          &  \pa \& \pb  &        \multicolumn{4}{c |}{$1$ generating pattern}   & $-$   \\ \cline{2-7}
                                           &  \pa \& \pd  &        \multicolumn{4}{c |}{$2$ generating patterns}    & $+$   \\ \bottomrule
\end{tabular}
}
\caption{Existing measures cannot yield desired relevance, unless we assert $\mathcal{M}_3$ (discipline) is always more selective than $\mathcal{M}_2$ (location), while PReP can achieve this by recognizing the co-occurrence of multiple generating patterns.}\label{tab::linearity}
\vspace{-18pt}
\end{table}

\vspace{12pt}

\vpara{The unified model.}
For notation convenience, we use the bold italic form to represent the corresponding matrix or vector of each symbol with subscripts.
For instance, the $(s, t)$ element of $\bmP$ is $P_{st}$ and the $t$-th element of $\bmeta$ is $\eta_t$.
Under this notation, combining Eq.~\eqref{eq::path_gen}, \eqref{eq::rho_prior}, and \eqref{eq::phi_prior}, with Eq.~\eqref{eq::decoupling} and \eqref{eq::gen_pat} substituted into Eq.~\eqref{eq::path_gen}, yields the total likelihood of the full PReP model 
\begin{align*}
\ttl 
&= p(\bm{P}, \bmeta, \bmrho, \bmPhi, \bmTheta \; | \; \alpha, \beta) \\
&= \left \{\prod_{u \in \nodeset} \gammadist{\rho_u \; ; \; (\alpha, 1)} \right \} \cdot \left \{ \prod_{s \in \pairset} \dirdist{K}{\bmphi_s \; ; \; \beta} \right \} \\
&\;\,\,\cdot \left \{ \prod_{\substack{s \in \pairset \\ (u, v)=s}} \prod_{t=1}^T \expdist{P_{st} \; ; \; \frac{\eta_t}{\rho_u \rho_v \sum_{k=1}^K \phi_{sk} \theta_{kt}}} \right \} \numberthis \label{eq::total_likelihood}
\end{align*}

\subsection{The PReP Relevance Measure}\label{sec::relevance}
Given the unified model (Eq.~\eqref{eq::total_likelihood}), we have two options to derive relevance measure using likelihood: 
(i) find the maximum a posteriori estimate for all parameters and compute the total likelihood of the observed data, and 
(ii) consider all model parameters as hidden variables and define the relevance as the marginal likelihood of the observed data. 
However, the marginal likelihood does not have a closed-form representation in our case, nor can we approximate it with regular Markov chain Monte Carlo algorithms due to the large number of hidden variables.
Therefore, we adopt the first option and defer the other to future work.

Once the model parameters $\{\bmeta, \bmrho, \bmPhi, \bmTheta\}$ are estimated, we define the PReP relevance for a node pair $s = (u, v)$ as the negative log-likelihood involving this node pair, $- \log p(\bmP_{s, :}, \bmphi_s \; | \; \bmTheta, \bmrho, \bmeta, \alpha, \beta)$, without the log term as in the derivation of PathSim in Sec.~\ref{sec::proba_view}
\begin{equation}\label{eq::relevance}
r(s) =  \sum_{t=1}^T \frac{\eta_t P_{st}}{\rho_u \rho_v  \sum_{k=1}^K \phi_{sk} \theta_{kt}} + (1 - \beta) \sum_{k=1}^K \log \phi_{sk}.
\end{equation}
Note that PathCount, PathSim, and JoinSim discussed in Sec.~\ref{sec::proba_view} are special cases of this PReP relevance measure, when $\{\bmeta, \bmrho, \bmPhi, \bmTheta\}$ are heuristically specified accordingly.

\section{Model Inference}
In this section, we introduce the inference algorithm for the PReP model (Eq.~\eqref{eq::total_likelihood}) proposed in Sec.~\ref{sec::model}.

\subsection{The Optimization Problem}
We find the maximum a posteriori (MAP) estimate for model parameters by minimizing the negative log-likelihood of the proposed model (Eq.~\ref{eq::total_likelihood}), which, with an offset of a constant, is given by

\begin{align*}
\obj 
&= \sum_{u \in \nodeset} (\rho_u -  (\alpha - 1) \log \rho_u) - (\beta - 1) \sum_{s \in \pairset} \sum_{k=1}^K \log \phi_{sk} \\
&+ T \sum_{(u, v) \in \pairset} (\log \rho_u + \log \rho_v) - |\pairset| \sum_{t=1}^T \log \eta_t \\
&+ \sum_{\substack{s \in \pairset \\ (u, v)=s}} \sum_{t=1}^T \left[ \log \sum_{k=1}^K \phi_{sk} \theta_{kt} + \frac{\eta_t P_{st}}{\rho_u \rho_v  \sum_{k=1}^K \phi_{sk} \theta_{kt}} \right],  \numberthis \label{eq::obj}
\end{align*}
and the optimization problem is therefore
\begin{equation}
\min_{\bmeta, \bmrho, \bmPhi, \bmTheta} \obj(\bmeta, \bmrho, \bmPhi, \bmTheta).
\end{equation}

We solve the above minimization problem with an iterative algorithm to be detailed in the following Sec.~\ref{sec::alg}.

\subsection{The Inference Algorithm}\label{sec::alg}
We iteratively update one of $\bmeta$, $\bmrho$, $\bmPhi$, and $\bmTheta$ when the others are fixed. The inference algorithm is summarized in Algorithm~\ref{alg::prep}.

\begin{algorithm}[t!]
\DontPrintSemicolon
\SetKwInOut{Input}{Input}
\SetKwInOut{Output}{Output}
\Input{the observed path counts $\bmP$ and the hyperparameters}
\Output{the model parameters $\bmeta$, $\bmrho$, $\bmPhi$, and $\bmTheta$}
\Begin{
  Initialize $\bmrho$, $\bmPhi$, and $\bmTheta$\;
  \While{not converged}{
    Update $\bmeta$ by the closed-form Eq.~\eqref{eq::update_mu}\;
    \While{not converged}{    
      \For{$u \in \nodeset$}{
        Update $\rho_u$ by the closed-form solution to Eq.~\eqref{eq::update_rho}\;
      }
    }
    Update $\bmPhi$ via parallelized PGD with gradient in Eq.~\eqref{eq::update_phi}\;
    Update $\bmTheta$ via PGD with gradient in Eq.~\eqref{eq::update_theta}\;
  }
}
\caption{Inference algorithm for the PReP model}\label{alg::prep}
\end{algorithm}

\vpara{Update $\bmeta$ given $\{ \bmrho, \bmPhi, \bmTheta \}$}. 
Once given $\bmrho$, $\bmPhi$, and $\bmTheta$, the optimal $\bmeta$ that minimizes $\obj$ in Eq.~\eqref{eq::obj} has a closed-form solution. 
One can derive this closed-form update formula by looking back to the total likelihood Eq.~\eqref{eq::total_likelihood}, since
\begin{align*}
\ttl
&\propto \prod_{s \in \pairset} \prod_{t=1}^T \expdist{P_{st} \; ; \; \frac{\eta_t}{\tau_s \sum_{k=1}^K \phi_{sk} \theta_{kt}}} \\
&= \prod_{t=1}^T \left[\expdist{\frac{1}{|\pairset|} \sum_{s \in \pairset} \frac{P_{st}}{\tau_s \sum_{k=1}^K \phi_{sk} \theta_{kt}} \; ; \; \eta_t} \right]^{|\pairset|},
\end{align*}
where $\tau_s = \rho_u \rho_v$ for node pair $s = (u, v)$.
Using the property of exponential distributions, we find the $\bmeta$ that maximizes $\ttl$, and hence minimizes $\obj$, can be computed by
\begin{equation}\label{eq::update_mu}
\eta_t = \left(\frac{1}{ |\pairset|} \sum_{s \in \pairset} \frac{P_{st}}{\tau_s \sum_{k=1}^K \phi_{sk} \theta_{kt}} \right)^{-1} .
\end{equation}

\vpara{Update $\bmrho$ given $\{ \bmeta, \bmPhi, \bmTheta \}$}.
Unlike $\bmeta$, closed-form formula for updating $\bmrho$ does not exist because (i) $\bmrho$ has an informative prior, and (ii) the generating process for paths between node pair $(u, v)$ involves the coupling of $\rho_u$ and $\rho_v$. Fortunately, the gamma distribution is the conjugate prior to the exponential distribution. 
Therefore, for each $u$, when the rest $\{\rho_v\}_{v \neq u}$ are fixed, the closed-form update formula for $\rho_u$ can be derived as follows. 
Denote $\{\xi_s\}_{s \in \pairset}$ the following quantities that are fixed during the $\bmrho$ update phase
$$
\xi_s \coloneqq \sum_{t=1}^T \frac{\eta_t P_{st}}{\sum_{k=1}^K \phi_{sk} \theta_{kt}},
$$
and we have
$
\frac{\partial \obj}{\partial \rho_u} = \sum_{\substack{v \in \nodeset \setminus \{u\} \\ s=(u, v)} } \left[ \sum_{t=1}^T \frac{1}{\rho_u} - \frac{\xi_s}{\rho_u^2 \rho_v} \right] - \frac{\alpha - 1}{\rho_u} + 1.
$
Setting this partial derivative to $0$ leads to
\begin{equation}\label{eq::update_rho}
\rho_u^2 + \left[ (|\nodeset|-1) \cdot T -(\alpha - 1) \right] \rho_u - \sum_{\substack{v \in \nodeset \setminus \{u\} \\ s=(u, v)} } \frac{\xi_s}{\rho_v} = 0. 
\end{equation}

Note that Eq.~\eqref{eq::update_rho} is a single-variable quadratic equation with one positive and one negative roots. Furthermore, $\obj$ is convex \wrt{} $\rho_u$ on the positive half-axis, and the positive root is a minimum of $\obj$. 
Therefore, the optimal $\rho_u$ that minimizes $\obj$ is given by the positive root of the quadratic equation (Eq.~\eqref{eq::update_rho}), which has closed-form solution. 
Holistically, we update $\bmrho$ by iterating through $u \in \nodeset$ to update $\rho_u$ with the aforementioned closed-form solution to Eq.~\eqref{eq::update_rho}. 

\vpara{Update $\bmTheta$ given $\{ \bmeta, \bmrho, \bmPhi \}$}.
To update $\bmTheta$, we use the projected gradient descent (PGD) algorithm \cite{nesterov2013introductory}. The gradient is given by
\begin{equation}\label{eq::update_theta}
\frac{\partial \obj}{\partial \bmTheta} = \bmPhi\trans \left[\frac{1}{\bmPhi \bmTheta} - \frac{\bmP}{(\bmtau (\bmeta^{\circ -1})\trans)\circ(\bmPhi \bmTheta)^{\circ 2}} \right],
\end{equation} 
where $[\cdot]\circ[\cdot]$, $\frac{[\cdot]}{[\cdot]}$, and $[\cdot]^{\circ [\cdot]}$ are element-wise multiplication, division, and power. 
Additional constraint fed into PGD is that each row of $\bmTheta$ lies in the standard $(T-1)$-simplex, \ie, $\sum_{t=1}^T \theta_{kt} = 1$ for all $k \in \{1, ..., K\}$ and $\theta_{kt} \geq 0$ for all $(k, t) \in \{1, ..., K\} \times \{1, ..., T\}$.
Projection onto the standard simplex or the direct product of multiple standard simplices can be achieved efficiently using the method introduced in \cite{duchi2008efficient}.

\vpara{Update $\bmPhi$ given $\{ \bmeta, \bmrho, \bmTheta \}$}.
Similarly, we use PGD to update $\bmPhi$, where the gradient is given by
\begin{equation}\label{eq::update_phi}
\frac{\partial \obj}{\partial \bmPhi} = \left[\frac{1}{\bmPhi \bmTheta} - \frac{\bmP}{(\bmtau (\bmeta^{\circ -1})\trans)\circ(\bmPhi \bmTheta)^{\circ 2}}\right] \bmTheta\trans - \frac{\beta - 1}{\bmPhi}.
\end{equation}
However, directly updating the entire $\bmPhi$ using PGD can be problematic, because the row number of $\bmPhi$ is the same as the number of nontrivial node pairs, $|\pairset|$, which can be significantly larger than that of $\bmTheta$. 

Fortunately, we can decompose the update scheme for $\bmPhi$ by rows, because each row is independent from the others.
Specifically, we update each row $s$ using PGD in parallel, with gradient 
$
\frac{\partial \obj}{\partial \bmPhi_{s, :}} = \left[\frac{1}{\bmPhi_{s, :} \bmTheta} - \frac{\bmP_{s, :}}{(\tau_s (\bmeta^{\circ -1})\trans)\circ(\bmPhi_{s, :} \bmTheta)^{\circ 2}} \right] \bmTheta\trans - \frac{\beta - 1}{\bmPhi_{s, :}},
$
and constraints $\sum_{k=1}^K \phi_{sk} = 1$ for all $s \in \pairset$ and $\phi_{st} \geq 0$ for all $(s, k) \in \pairset \times \{1, ..., K\}$.

\subsection{Implementation Details}\label{sec::implementation}
For program reproducibility, we provide details in parameter initialization and computational singularity handling.

Since the inference algorithm starts with updating $\bmeta$, no initialization for $\bmeta$ is needed. 
$\bmrho$ is initialized by drawing random samples from its prior distribution, $ \gammadist{\alpha, 1}$, where $\alpha$ is estimated from data as discussed in Sec.~\ref{sec::model}.
$\bmPhi$ is initialized uniformly at random within the row-wise simplex constraint. For $\bmTheta$, the first $T$ rows of this $K \times T$ matrix are initialized to be an identity matrix, because many node pairs with paths in between involve only one meta-path, and we initialize the rest $K - T$ rows uniformly at random within the row-wise simplex constraint. This choice is out of the consideration that the PReP model is not convex over all parameters.

Dirichlet distribution is defined over open sets with unbounded probability density function.
As a result, when using MAP, certain components of $\bmPhi$ can be inferred to approach the singularities along the boundary. 
Therefore, in practice, we let $\bmPhi$ to be bounded away from the boundary with an infinitesimal quantity $\delta$, \ie, each of its entries must not only be positive, but also be greater or equal to $\delta$.
In this way, we keep the capability of Dirichlet distribution in modeling \crosssyn, while ensuring the model is computationally meaningful.
In our experiment, we set $\delta = 10^{-50}$.
With this constraint, the domain of definition for $\bmPhi$ is no longer a standard simplex as discussed in \cite{duchi2008efficient}.
For this reason, we provide the algorithm for efficient projection onto this shrunken simplex, $\{\mb{x} \in \mathbb{R}^K | x_i \geq \delta, \sum_{i=1}^K x_i = 1\}$ in the Appendix, which is required by the inference algorithm.
Note that if one wishes to evade the point estimation of parameters in the PReP model, Eq.~\eqref{eq::total_likelihood}, and thereby avoid computational singularity, they can treat all model parameters as hidden variables and derive relevance from the marginal likelihood of the observed data as discussed in Sec.~\ref{sec::relevance}.
The exploration of this direction requires novel method, such as a sampling algorithm design for our model, to efficiently calculate marginal likelihood, and we defer this to future work.

\section{Experiments}
In this section, we quantitatively evaluate the proposed model on two publicly available real-world HINs: Facebook and DBLP.
We first describe the datasets and the unsupervised tasks used for evaluation.
Baselines and model variations for comparison are then introduced.
Afterward, we present experiment results together with discussions,
which demonstrate the advantage of using probability as the backbone of relevance. 

\subsection{Data Description and Evaluation Tasks}
In this section, we introduce the two publicly available real-world datasets and the evaluation tasks.

\vpara{The Facebook dataset.}
This dataset \cite{mcauley2012learning} contains nodes of $11$ types, including user, major, degree, school, hometown, surname, location, employer, work-location, work-project, and other. It consists of $5,621$ nodes and $98,023$ edges, among which $4,167$ nodes are of the user type.
We aim to determine the relevance between users, using $10$ meta-paths, each of the form [user]--[X]--[user], where X is any of the above $11$ node types except for other.

To derive ground truth label between user pairs for evaluation, we use being friends on Facebook as a proxy for being relevant.
This dataset is collected by recruiting participants to label their own Facebook friends
It consists of $10$ distinct ego networks, where an ego network consists of one ego user and all her friends together with edges attached to these users.
We hence perform one sub-task for each ego network, where the compared measures are used to calculate the relevance between all pairs of non-ego users in this ego network.

We use two evaluation metrics widely adopted for tasks with multiple relevant instances: the area under the receiver operating characteristic curve (ROC-AUC) and the area under precision-recall curve (AUPRC).
The receiver operating characteristic curve (ROC) is created by plotting true positive rate against false positive rate as the threshold varies, while the precision-recall curve (PRC) is drawn by plotting precision against recall as the threshold varies. 
Higher values are more preferred for both ROC-AUC and AUPRC.
We further average each of the above metrics across ego networks with the following methods -- uni.: averaging over all ego networks uniformly; rel.: weighting by the number of relevant pairs in each ego network; tot.: weighting by the total number of pairs in each ego network.

\vpara{The DBLP dataset.}
This dataset is derived from the DBLP dataset processed by Tang et al. \cite{tang2012unified} containing computer science research papers together with author names and publication venue associated to each paper. It consists of $13,697$ nodes and $19,665$ edges, among which $1,546$ nodes are of the author type.
Notably, in this dataset, the same author name associated with two papers may not necessarily be the same person. 
Based on this fact, we design an entity resolution task as follows.
First, we use the labels made available by Tang et al. \cite{tang2012unified} to group all author name mentions corresponding to one person to define an author node. In this way, an author node is linked to multiple papers written by her.
Then, for each author name, we split the author node with the most author name mentions into two nodes, and we define two nodes to be relevant if and only if they actually refer to the same person.
Finally, we perform one sub-task for each author name, where the compared measures are used to calculate the relevance between all pairs of nodes with the same author name.

We use $14$ meta-paths in this task, each of the form [author]--[paper]--[venue domain]--[paper]--[author], where a node of the venue domain type corresponds to one of the $14$ computer science research areas. 
The definition of the $14$ areas is derived from the Wikipedia page: {List of computer science conferences}\footnote{https://en.wikipedia.org/wiki/List\_of\_computer\_science\_conferences}.
Since only one relevant pair exists in each sub-task, the mean reciprocal rank (MRR) is used as the evaluation metric, where, for each sub-task, the reciprocal rank is the reciprocal of the rank of the relevant pair.
Higher values indicate better results for MRR.
We also average the above metrics across different sub-tasks using three methods: uni., rel., and tot. Note that uni. and rel. are equivalent in this entity resolution task because each sub-task has exactly one relevant pair.

\begin{table*}[t!]
\centering
\resizebox{\textwidth}{!}{
\begin{tabular}{ l | c | c || c | c | c | c | c | c | c | c || c | c | c || c }
\toprule \hline
\multirow{2}{*}{Dataset} & \multicolumn{2}{ c ||}{\multirow{2}{*}{Metric}} & \multicolumn{2}{ c |}{PathCount} & \multicolumn{2}{ c |}{PathSim} & \multicolumn{2}{ c |}{JoinSim} & \multicolumn{2}{ c ||}{SimRank} & \multicolumn{4}{ c }{PReP}  \\ \cline{4-15}
 & \multicolumn{2}{ c ||}{} & Mean & SD & Mean & SD & Mean & SD & Mean & SD & No-NV & No-PS & No-CS & (full) \\ \hline
\multirow{6}{*}{Facebook}  & \multirow{3}{*}{ROC-AUC}   & uni. 	& 0.8056 & 0.8598 & 0.8367 & 0.8586 & 0.8326 & 0.8547 & 0.7977 & 0.8303 & 0.8310 & 0.6702 & 0.8689 & \textbf{0.8850} \\ \cline{3-15}
                                          &                                           & rel.	& 0.8612 & 0.8879 & 0.8578 & 0.8888 & 0.8556 & 0.8872 & 0.8076 & 0.8596 & 0.8556 & 0.6713 & 0.8880 & \textbf{0.9133} \\ \cline{3-15}
                                          &                                           & tot.	& 0.8558 & 0.8849 & 0.8577 & 0.8866 & 0.8557 & 0.8851 & 0.8096 & 0.8594 & 0.8547 & 0.6773 & 0.8893 & \textbf{0.9139} \\ \cline{2-15}
                                          & \multirow{3}{*}{AUPRC}      & uni.	& 0.2456 & 0.2832 & 0.2370 & 0.2845 & 0.2340 & 0.2803 & 0.2055 & 0.2435 & 0.2183 & 0.1650 & \textbf{0.3273} & \textbf{0.3269} \\ \cline{3-15}
                                          &                                           & rel.	& 0.2496 & 0.3048 & 0.2142 & 0.2873 & 0.2117 & 0.2837 & 0.1764 & 0.2408 & 0.2067 & 0.1283 & 0.3354 & \textbf{0.3486} \\ \cline{3-15}
                                          &                                           & tot.	& 0.2107 & 0.2542 & 0.1841 & 0.2460 & 0.1821 & 0.2432 & 0.1523 & 0.2071 & 0.1760 & 0.1089 & 0.3010 & \textbf{0.3080} \\ \hline
\multirow{2}{*}{DBLP}         & \multirow{2}{*}{MRR}   & uni./rel.	& 0.8091 & 0.8130 & 0.6922 & 0.7003 & 0.7454 & 0.7538 & 0.6636 & 0.6738 & 0.8223 & 0.8494 & 0.8365 & \textbf{0.8517} \\ \cline{3-15}
                                          &                                           & tot.	& 0.7839 & 0.7871 & 0.6612 & 0.6731 & 0.7128 & 0.7244 & 0.6302 & 0.6357 & 0.8234 & \textbf{0.8407} & 0.8264 & 0.8391 \\ \hline
\bottomrule
\end{tabular}
}
\caption{Quantitative evaluation results on two real-world datasets using the proposed measure, PReP, and other measures.}\label{tab::quant}
\vspace{-12pt}
\end{table*}

\subsection{Baselines and Variations}
In this section, we describe the meta-path-based baseline methods and variations of the PReP model, which are used to compare with our proposed full PReP model. 
Existing meta-path-based unsupervised HIN measures define relevance computation method on each meta-path and then use linear combination to find the composite score.
Therefore, each baseline consists of two parts: (i) the base measure that calculates the relevance score on one meta-path, and (ii) the weights assigned to different meta-paths used in the linear combination.
The $4$ base measures we used are:
\begin{itemize}
\item
\textbf{PathCount} \cite{sun2011pathsim}. 
$\mathit{PathCount}_{\mb{w}}(s) \coloneqq \sum_t w_t P_{st}$.
\item
\textbf{PathSim} \cite{sun2011pathsim}. 
$\mathit{PathSim}_{\mb{w}}(s) \coloneqq \sum_t w_t \cdot \frac{2 \cdot P_{\langle uv \rangle t}}{P_{\langle uu \rangle t} + P_{\langle vv \rangle t}}$.
\item
\textbf{JoinSim} \cite{xiong2015top}. 
$\mathit{JoinSim}_{\mb{w}}(s) \coloneqq \sum_t w_t \cdot \frac{P_{\langle uv \rangle t}}{\sqrt{P_{\langle uu \rangle t} \cdot P_{\langle vv \rangle t}}}$.
\item
\textbf{SimRank}. We adopt SimRank \cite{jeh2002simrank} with meta-path constraints. Let $\bm{A}$ be a matrix, where $A_{u v}$ is the number of paths under this meta-path between node pair $(u, v)$ after column normalization. The SimRank score is then given by $S_{u v}$, where $\bm{S}$ is the solution to
$\bm{S} = \max \{C \cdot (\bm{A}\trans \bm{S} \bm{A}), \bm{I} \}$,
and $C$ is the decay factor to be specified. Note that we use SimRank instead of random walk with restart because SimRank is a symmetric relevance measure.
\end{itemize}
Without any supervision available, we use 2 heuristics to determine the weights $\mb{w}$ for linear combination.
\begin{itemize}
\item
\textbf{Mean}. Let $w_t$ be the reciprocal of the mean of all scores computed using the corresponding base measure on meta-path $t$.
\item
\textbf{SD}. Let $w_t$ be the reciprocal of the standard deviation of all scores computed using the corresponding base measure on meta-path $t$. Note that this heuristic normalized the original score in the way similar to $z$-score.
\end{itemize}
Combining the aforementioned $4$ base measures and $2$ heuristic for setting weights, we have $8$ baselines in total. 

Additionally, we also experiment with three variations of PReP, which are partial models with one of the three components knocked out from the full PReP model.
\begin{itemize}
\item
No \nodevis (No-NV): Set $\bmrho = \bm{1}_{|\nodeset|}$, and do not update $\bmrho$ during model inference.
\item
No \pathsel (No-PS): Set $\bmeta = \bm{1}_T$, and do not update $\bmeta$ during model inference.
\item
No \crosssyn (No-CS): Set $\bmPhi = {\bm{1}_{|\nodeset| \times K}} / {K} $, $\bmTheta = {\bm{1}_{|\nodeset| \times T}}/{T}$, and do not update $\bmPhi$ and $\bmTheta$ during model inference.
\end{itemize}
Note that $\bm{1}_M$ stands for all one column vector of size $M$ and $\bm{1}_{M \times N}$ denotes all one matrix of size $M \times N$.

\subsection{Effectiveness and Discussion}

In this section, we present the quantitative evaluation results on both the Facebook and the DBLP datasets. 
We tune the decay factor $C$ in the baseline measure, SimRank, to have the best performance with $C = 0.5$ for both SimRank-Mean and SimRank-SD on Facebook, and $C = 0.8$ for SimRank-Mean, $C = 0.7$ for SimRank-SD on DBLP. 
We set hyperparameters of PReP as $K = 15$ and $\beta = 10^{-4}$ for Facebook and $K = 14$ and $\beta = 10^{-2}$ for DBLP.
The choice of hyperparameters will be further discussed in this section.

As presented in Tab.~\ref{tab::quant}, PReP outperformed all $8$ baselines under various metrics. 
Moreover, PReP outperformed its $3$ variations under most metrics, suggesting each component of the model generally has a positive effect on the performance of the full PReP model.
Note that under MRR (tot.), PReP performed slightly worse than PReP-No-PS, the partial model without $\eta_t$ for \pathsel.
This happened because, as discussed in Sec.~\ref{sec::model}, we cannot enforce task-specific design on \pathsel $\eta_t$ due to the lack of supervision, and we expect \pathsel $\eta_t$ to play a more important role in future work where relevance labels are provided as supervision.

Additionally, we have made the following observations.

\vpara{Heuristic methods cannot yield robust relevance measures.}
Compared with PathCount, both PathSim and JoinSim further model \nodevis, which penalizes the relevance with nodes that are highly visible. 
However, as Tab.~\ref{tab::quant} presents, PathSim and JoinSim cannot always outperform PathCount.
Moreover, JoinSim performs better than PathSim on DBLP, while PathSim is slightly better than JoinSim on Facebook.
We interpret these results as, PathSim and JoinSim model \nodevis in a deterministic heuristic way.
Unlike our generative-model-based measure that derives relevance measure based on parameters inferred from each HIN, the heuristic approaches adopted by PathSim and JoinSim have varying performance on different HINs.
This suggests being data-driven is a favorable property of PReP.

\vpara{Non-one-hot generating patterns help only when meta-paths correlate.}
In our experiment, we set $K = 14 = T$ for DBLP. 
Recall that we initialized the first $T$ rows of $\bmTheta$, the matrix representing the $K$ generating patterns, to be $T$ one-hot vectors corresponding to $T$ meta-paths.
We observed in the DBLP experiment that after model fitting, $\bmTheta$ was still the same as its initialization, meaning each inferred generating pattern only generated path instances under exactly one meta-path.
Moreover, by increasing the value of $K$, we did not see improvement in performance.
This observation is inline with the situation that it is not frequently seen that two authors both publish papers in two distinct research areas, where the $14$ areas on the Wikipedia page have been defined to be distinct areas including theory, software, parallel computing, \etc{} 
In this case, it is preferred to model synergy across every pair of meta-paths, and not to employ any non-one-hot generating patterns.

On the other hand, we used $K = 15 > T$ for Facebook, and we did observe non-one-hot generating patterns after model fitting.
The most popular non-one-hot generating pattern consisted of three meta-paths: [user]--[hometown]--[user], [user]--[school]--[user], and [user]--[user]--[user], where we define popularity of a generating pattern as the fraction of node pairs adopting this pattern, \ie, $\mathit{pop}(k) = \sum_{s \in \pairset} \phi_{sk}$.
This generating pattern corresponds to two users sharing the same hometown, the same school, and having common friends.
This scenario is common for two people sharing similar friend group back in the hometown school.

\begin{figure}[t]
  \centering
  \begin{subfigure}[m]{0.49\linewidth}
    \centering\includegraphics[width=\linewidth]{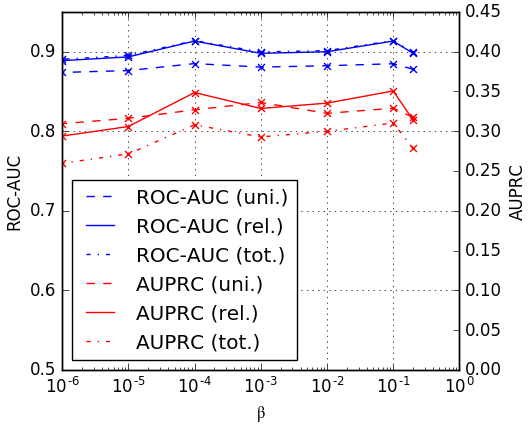}
    \vspace{-12pt}
    \caption{Facebook}\label{fig::param_sens_a}
    \vspace{-6pt}
  \end{subfigure}
  $\;$
  \begin{subfigure}[m]{0.45\linewidth}
    \centering\includegraphics[width=\linewidth]{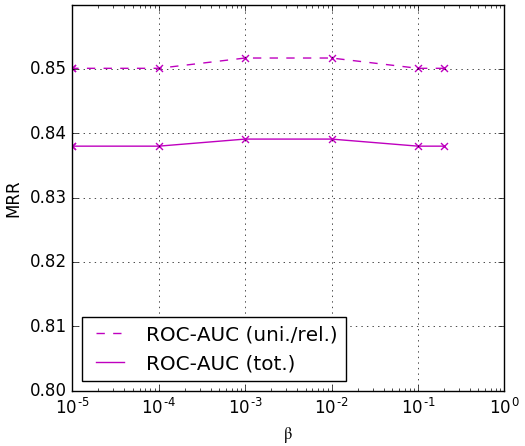}
    \vspace{-12pt}
    \caption{DBLP}\label{fig::param_sens_b}
    \vspace{-6pt}
  \end{subfigure}
  \caption{Performance with varying $\beta$.}\label{fig::param_sens}
  \vspace{-12pt}
\end{figure}

\vpara{Sensitivity of $\beta$ in modeling \crosssyn.}
In the PReP model (Eq.~\eqref{eq::total_likelihood}) and relevance measure (Eq.~\eqref{eq::relevance}), the concentration parameter $\beta$ of the Dirichlet prior controls the extent to which we boost \crosssyn. Experiment results in Fig~\ref{fig::param_sens} shows performance of PReP do not significantly change around the values we have set for $\beta$, \ie, $10^{-4}$ for Facebook and $10^{-2}$ for DBLP.

\section{Related Work}
In this section, we review the study on HIN relevance.
The problem of deriving relevance between node pairs has been extensively studied for homogeneous information networks. 
Relevance measures of this type include the random walk based \textit{Personalized PageRank} and \textit{SimRank} \cite{jeh2002simrank}, the neighbor-based \textit{common neighbors} and \textit{Jaccard's coefficient}, the path-based \textit{Katz} \cite{han2011data}, \etc{} 
To generalize relevance from the homogeneous networks to the typed heterogeneous case, researchers have been exploring from multiple perspectives.
One perspective, as in \textit{PathCount} and \textit{PathSim} from \cite{sun2011pathsim} and \textit{Path-Constrained Random Walk} from \cite{lao2010relational}, is to first compute relevance score along each meta-path, and then glue scores from all types together via linear combination to establish the composite measure. 
A great many applications \cite{kuck2015query, shi2017survey, sun2013mining, yu2014personalized,zhuang2014mining} based on this meta-path paradigm with linear combination have been proposed. 
Our proposed method follows this meta-path paradigm, but goes beyond linear combination to model \crosssyn that we have observed from real-world HINs.
Another perspective is to go beyond meta-path and derive relevance based on the more complex graph structures \cite{fang2016semantic, huang2016meta}. 
While these approaches can yield good performance, they differ from our proposed methods for further entailing label information or expertise in designing graph structure. 
Also, they do not carry probabilistic interpretations.
Besides, people have explored the idea of incorporating richer information \cite{he2014exploiting, yao2014pathsimext} to define more effective relevance scoring functions, or adding supervision to derive task-specific relevance measures \cite{chen2017task, wang2011learning, yu2012user}.
While being valuable, these works are out of the scope of the problem we study in this paper, where we address the basic, unsupervised case with no additional information as our starting point of studying HIN relevance from the probabilistic perspective.

\section{Conclusion and Future Work}
Inspired by the probabilistic interpretation of existing path-based relevance measures, we studied HIN relevance from a probabilistic perspective.
We identified \crosssyn as one of the three characteristics that we deem important for HIN relevance.
A generative model was proposed to derive a novel path-based relevance measure, PReP, which could capture the three important characteristics.
An inference algorithm was also developed to find the maximum a posteriori (MAP) estimate of the model parameters, which entailed non-trivial tricks.
Experiments on real-world HINs demonstrated the effectiveness of our relevance measure, which is data-driven and tailored for each HIN.

Future work includes the exploration of defining relevance from the proposed PReP model with marginal likelihood as discussed in Sec.~\ref{sec::relevance}.
Further add-on designs to adapt the proposed model to a supervised setting are also worth exploring to unleash the potential of our model.

\vpara{Acknowledgments.}
We thank our colleagues and friends for the enlightening discussions: Jason Jian Ge, Jiasen Yang, Carl Ji Yang, and many members of the Data Mining Group at UIUC.
We also thank the anonymous reviewers for their insightful comments. 
This work was sponsored in part by the U.S. Army Research Lab. under Cooperative Agreement No. W911NF-09-2-0053 (NSCTA), National Science Foundation IIS-1320617 and IIS 16-18481, and grant 1U54GM114838 awarded by NIGMS through funds provided by the trans-NIH Big Data to Knowledge (BD2K) initiative (www.bd2k.nih.gov). The views and conclusions contained in this document are those of the author(s) and should not be interpreted as representing the official policies of the U.S. Army Research Laboratory or the U.S. Government. The U.S. Government is authorized to reproduce and distribute reprints for Government purposes notwithstanding any copyright notation hereon.


\section*{Appendix}
We provide the algorithm for efficient projection onto the standard simplex shrunk by $\delta$, $\{\mb{x} \in \mathbb{R}^K | x_i \geq \delta, \sum_{i=1}^K x_i = 1\}$, in Algorithm~\ref{alg::simplex}.

\begin{algorithm}[h!]
\DontPrintSemicolon
\SetKwInOut{Input}{Input}
\SetKwInOut{Output}{Output}
\Input{the original vector $\mb{z} \in \mathbb{R}^K$ and the shrinking factor $\delta$}
\Output{the projection $\mb{x} \in \mathbb{R}^K$}
\Begin{
  Sort $\mb{z}$ into $\mb{u}$: $u_1 \geq u_2 \geq \ldots \geq u_K$\;
  $\rho \leftarrow \max \{1 \leq j \leq K | u_j + \frac{1}{j}(1 - \delta K - \sum_{i=1}^j u_i) > 0\}$\;
  $\lambda \leftarrow \frac{1}{\rho}(1 - \delta K - \sum_{i=1}^\rho u_i)$\;
  $x_i \leftarrow \max \{z_i + \lambda, 0\} + \delta$\;
}
\caption{Efficient projection onto shrunk simplex}\label{alg::simplex}
\end{algorithm}

The validity of this algorithm can be established in a way similar to the proof of the algorithm for standard simplex \cite{duchi2008efficient}.

\newpage

\bibliographystyle{ACM-Reference-Format}
\bibliography{sigproc}

\end{document}